\documentclass[
    ,final            % use final for the camera ready runs
  ]
  {aipproc}

\layoutstyle{6x9}

\def\gr{$\gamma$-ray\ }
\def\grs{$\gamma$-rays\ }

\begin{document}

\title{CTA and cosmic-ray diffusion in molecular clouds}

\classification{96.50.S-, 95.85.Pw, 98.38.Dq, 98.38.Mz}
\keywords      {cosmic rays, gamma-ray astronomy, interstellar molecular clouds, particle diffusion}

\author{G.~Pedaletti}{
  address={Institut de Ci\`encies de l'Espai (IEEC-CSIC),
              Campus UAB,  Torre C5, 2a planta,
              08193 Barcelona, Spain}
}

\author{D. F. Torres}{
  address={Institut de Ci\`encies de l'Espai (IEEC-CSIC),
              Campus UAB,  Torre C5, 2a planta,
              08193 Barcelona, Spain}
,altaddress={Instituci\'o Catalana de Recerca i Estudis Avan\c{c}ats (ICREA)} % additional visiting address
}

\author{S. Gabici}{
  address={Astroparticule et Cosmologie (APC), CNRS, Universit\'{e} Paris 7 Denis Diderot, Paris, France}
}
 
\author{E. de O\~{n}a Wilhelmi}{
  address={Max-Planck-Institut f\"{u}r Kernphysik (MPIK), P.O. Box 103980, 69029 Heidelberg, Germany}
, altaddress={Institut de Ci\`encies de l'Espai (IEEC-CSIC),
              Campus UAB,  Torre C5, 2a planta,
              08193 Barcelona, Spain}
}

\author{D. Mazin}{
  address={IFAE, Edifici Cn., Campus UAB, E-08193 Bellaterra, Spain}
}

\author{V. Stamatescu}{
  address={IFAE, Edifici Cn., Campus UAB, E-08193 Bellaterra, Spain}
}

\begin{abstract}
Molecular clouds act as primary targets for cosmic-ray interactions and are expected to shine in gamma-rays as a by-product of these interactions. Indeed several detected gamma-ray sources both in HE and VHE gamma-rays (HE: 100 MeV < E < 100 GeV; VHE: E > 100 GeV) have been directly or indirectly associated with molecular clouds. Information on the local diffusion coefficient and the local cosmic-ray population can be deduced from the observed gamma-ray signals. In this work we concentrate on the capability of the forthcoming Cherenkov Telescope Array Observatory (CTA) to provide such measurements. We investigate the expected emission from clouds hosting an accelerator, exploring the parameter space for different modes of acceleration, age of the source, cloud density profile, and cosmic ray diffusion coefficient. We present some of the most interesting cases for CTA regarding this science topic. The simulated gamma-ray fluxes depend strongly on the input parameters. In some cases, from CTA data it will be possible to constrain both the properties of the accelerator and the propagation mode of cosmic rays in the cloud.
\end{abstract}

\maketitle

%%%%%%%%%%%%%%%%%%%%%%%%%%%%%%%%%%%%%%%%%%%%
%% MAINMATTER
%%%%%%%%%%%%%%%%%%%%%%%%%%%%%%%%%%%%%%%%%%%%
\section{Introduction}
\textbf{The physical scenario} Emission in HE-VHE \grs is expected in spatial
coincidence with molecular clouds, resulting from the hadronic
interaction between cosmic-ray (CR) particles and the dense material
in the cloud acting as a target. Indeed, some MCs have been detected
in \grs in both the GeV and TeV domain \citep[see, e.g., ][]{ic443_magic,w28_hess,mc_fermi}. Moreover, it has been suggested that some of the still unidentified \gr sources might also be MCs illuminated by CRs that escaped from an accelerator located inside the cloud or in its proximity \citep{montmerle79,aha_ato,gabici2007,rodriguez2008}. In such cases the modeling of the emission involves the parametrization of the diffusion of charged particles. 
The study of this emission is extremely useful in unveiling the physics of CR sources. A more detailed study following the lines of the present contribution can be found in \cite{longer}.

If a power-law energy spectrum ($J_\mathrm{p}(E_\mathrm{p})=K E_\mathrm{p}^{-\gamma}$) is assumed for the intensity of primary CRs, the resulting \gr spectrum due to hadronic interactions would also follow a power-law spectrum ($F(E) \propto E^{-\Gamma}$). The \gr spectrum reproduces the spectrum of the parent particles, at high energies. Although, if we consider energy-dependent diffusion coefficient, the CR spectrum may differ from a simple power-law near the acceleration site. Here, we assumed that the source is point-like and located at the origin of the coordinate system. 
The acceleration and diffusion processes are computed following the approach of \cite{aha_ato}, with the appropriate solution of the diffusion-loss equation. The diffusion coefficient is assumed to depend on the CR energy only, as: $D(E)=D_\mathrm{10} (E\, / \, 10\,\textrm{GeV})^{\delta}$ cm$^{2}$ s$^{-1}$. 
We investigate the case of an accelerator at the center of a molecular cloud.

The resulting flux in \grs is mainly dependent on: diffusion coefficient and its energy dependence ($D_{10}, \delta$); age of the accelerator; type of accelerator (impulsive/ continuous); spectrum of injection ($\gamma$); fraction of energy in input (total energy in form of cosmic rays $\mathrm{W_p}=\eta10^{50}$ erg). An impulsive source of particles corresponds to the case when the bulk of relativistic cosmic-rays are released during times much smaller than the age of the accelerator itself. When the timescales are comparable, the source is referred to as a continuous injector. The total energy input is taken as $\mathrm{W_p}$ in the impulsive case, while the energy injection rate in the continuous case is of $L_\mathrm{p}=10^{37}$ erg s$^{-1}$, resulting in the same total input for accelerators with $\eta=1$ and with an age of a few hundreds of thousand years.
%All the parameters are in principle free, however, we mainly concentrate, as an example, on the case were the injection slope is $\gamma=2.2$ and the diffusion coefficient dependence on energy is $\delta=0.5$. This pair satisfies the observed CR spectrum data. Indeed, with these values, the index of the equilibrium spectrum in the galaxy is expected to be $\gamma+\delta=2.7$. 
%The typical value of the diffusion coefficient in the galaxy is $D_{10}=10^{28}$ cm$^{2}$ s$^{-1}$ \citep{ber_book}. However, this value is very uncertain and depends on the level
%of the magnetic turbulence in which particles propagate. 
In this scenario, a great variety of \gr spectra is expected. This can produce a variety of different GeV--TeV connections, some of which could explain the observed phenomenology \citep[see, e.g.,][]{tam_gevtev}. 

\textbf{CTA}
The Cherenkov Telescope Array (CTA) is an international project for the development of the next generation ground-based  very high energy (VHE) gamma-ray instrument.  
CTA will significantly advance with respect to the present generation
IACTs: it will feature an order of magnitude improvement in
sensitivity at the core energy range of 1 TeV, improve in its angular
and energy resolution, and provide wider energy coverage, see \cite{dc2010}. Indeed, the array is expected to have an unprecedented sensitivity down to $\sim$50 GeV and above $\sim$50 TeV, establishing a strong link with the satellite-based operations at low energies, namely the Large Area Telescope on board the {\it Fermi} satellite, see \cite{atwood2009} and water Cherenkov experiments at the highest energies \citep[e.g, HAWC, see][]{hawc2010}. Both a southern and a northern hemisphere observatory are foreseen. 

Detailed response functions for a proposed CTA array can be found in \cite{apissue_mc}. We use the procedure detailed there to calculate the spectral points and profiles from CTA simulated observations.

\section{CTA response}
\begin{figure}[!ht]
%\begin{center}
%\includegraphics[width=0.49\textwidth]{eps/enbreak_70gev1.eps}
\includegraphics[width=0.45\textwidth]{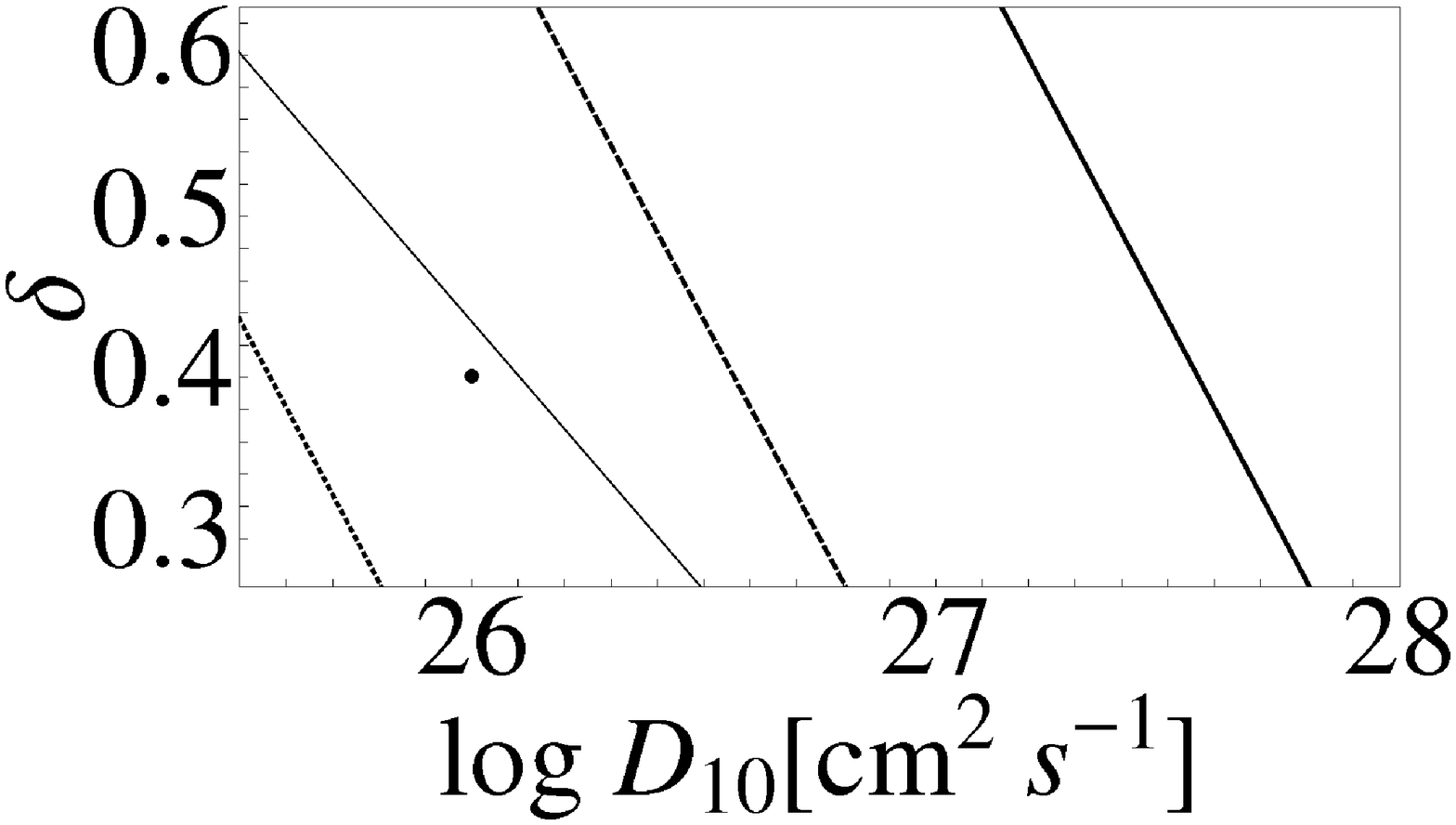}
\includegraphics[width=0.49\textwidth]{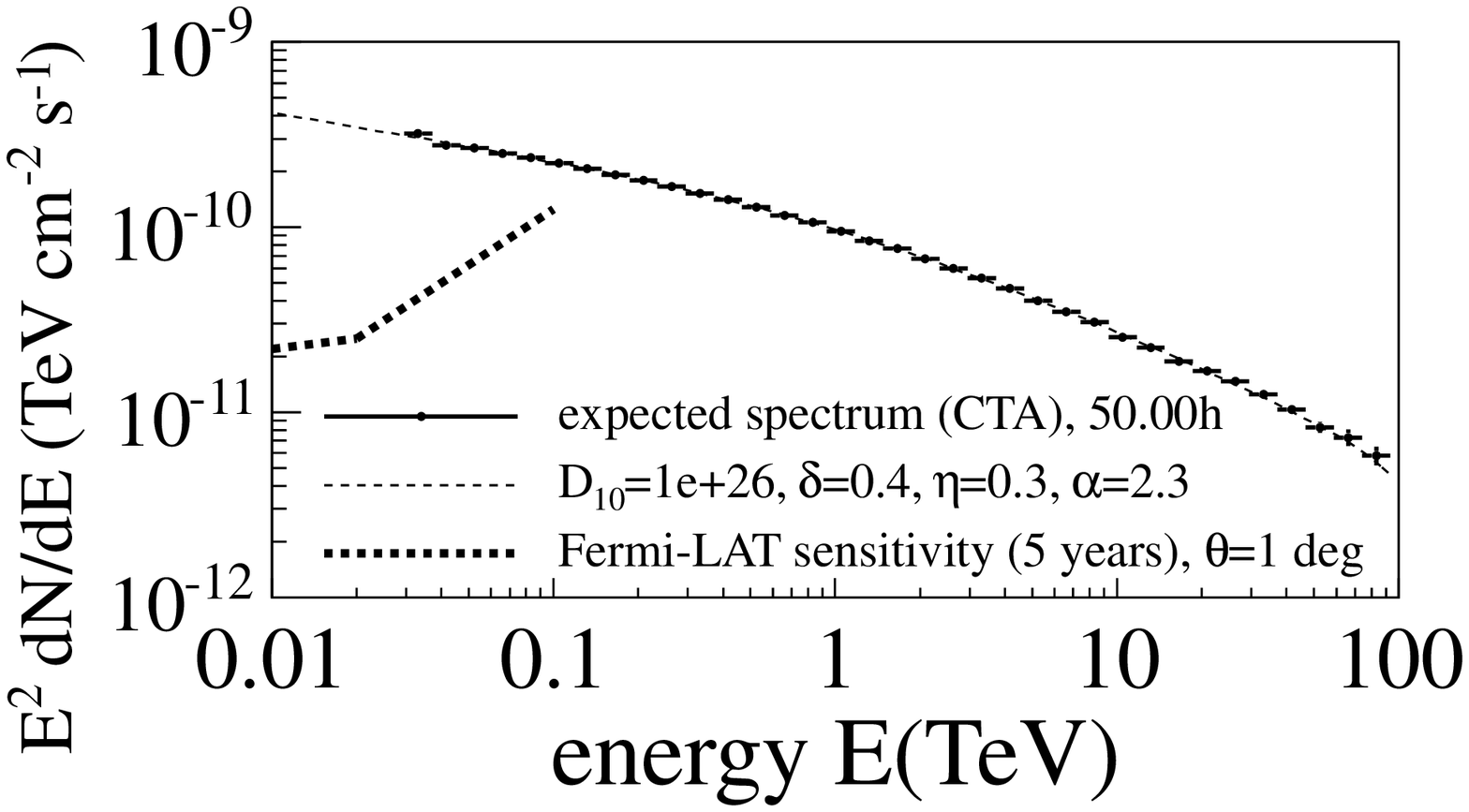}
\caption{\textbf{Left} Boundaries for $E_\mathrm{break}>70$ GeV in \gr (left from the boundary) for different ages of the accelerator ($10^{3,4,5}$ years are
represented by the tick lines, solid, dashed, dotted curves, respectively). The thin solid line represents the boundary for an age of $10^{4}$ years and $E_\mathrm{break}>1$ TeV. The point represents the parameters of the spectrum given in the right panel. \textbf{Right} CTA expected performances on the reconstruction of the intrinsic model for $D_{10}=10^{26}$ cm$^2$ s$^{-1}$, $\delta=0.4$, $\gamma=2.3$, $\eta=1/3$. The points are the expected spectral points from 50 hours of CTA observation time. The intrinsic spectrum is the only accepted model (see text). The expected 5 year point source sensitivity for Fermi/LAT is also given. This is calculated from the 1 year sensitivity in \citep{atwood2009}, linearly scaled with time at high energies ($>10$ GeV).} 
 \label{fig:break_chi}
%\end{center}
\end{figure}

\textbf{Spectral features} A constraint on the diffusion coefficient can come from the identification of a break in the \gr spectrum integrated from the entire cloud region. The break can be related to the minimum energy that can diffuse in the entire cloud over a timescale comparable to the age of the accelerator. Equating the diffusion timescale, with the age of the accelerator, one obtains the constraints shown in Fig. \ref{fig:break_chi}, left, corresponding to:
\begin{equation}\label{eq:e_break}
 E_\mathrm{break}=10\left(\frac{R_\mathrm{cloud}^2}{6 D_\mathrm{10} t_\mathrm{age}}\right)^{1/\delta} \textrm{GeV}.
\end{equation}
For scenarios with galactic average diffusion ($D_{10}=10^{28}$ cm$^2$ s$^{-1}$) and energy dependence parameter in the range $\delta=[0.3 .. 0.6]$, the corresponding break in \gr emission will always be at energies below the CTA energy acceptance. This is accurate for impulsive accelerators. In the continuous acceleration case, new injections of high energy particles will smooth the effect of a break.
At energies higher than the break given by Eq. \ref{eq:e_break}, the particle spectrum will follow a power-law form composed by the slope of the injection spectrum and the energy dependence of the diffusion coefficient (i.e $\gamma+\delta$). Therefore the \gr emission will also show a power-law behavior, reducing the capability of constraining the parameter space from \gr data.

Let us assume that a molecular cloud with measured mass and distance is detected at
TeV energies. Let us further assume that from multiwavelength observations we
identified a possible accelerator of CRs responsible for the gamma ray emission
and that an estimate of the age of that accelerator is known. In Fig. \ref{fig:break_chi} we show the
simulated spectra for the gamma ray emission for such an accelerator, which is
assumed to be impulsive and with an age of $10^4$ years.
The molecular cloud is assumed to have a mass of $M_\mathrm{5}=10^5 M_\odot$ and a radius of 20 pc (hence with an average density of $n_\mathrm{H}=130$ cm$^{-3}$) located at a distance of $d=$1 kpc.
The other parameters are varied on a discretized grid and the corresponding observed spectrum is simulated from CTA responses.
The best fitting model belonging to the parameter space grid is found along with the models with a $\chi^2/dof$ not more different than unity from the best fit model. The right panel of Fig. \ref{fig:break_chi} shows the case of $D_\mathrm{10}=10^{26}$ cm$^2$ s$^{-1}$, $\delta=0.4$, $\gamma=2.3$, $\eta=1/3$. Thanks to the high flux reached in this case and the presence of a break, the spectrum is reconstructed easily to the intrinsic parameters, with a break at $E_\gamma\approx2$ TeV and slopes $\Gamma_\mathrm{1}=2.3$ and $\Gamma_\mathrm{2}=2.7$ below and above the break, respectively.

\textbf{Morphology}
\begin{figure}[!ht]
\includegraphics[width=0.45\textwidth]{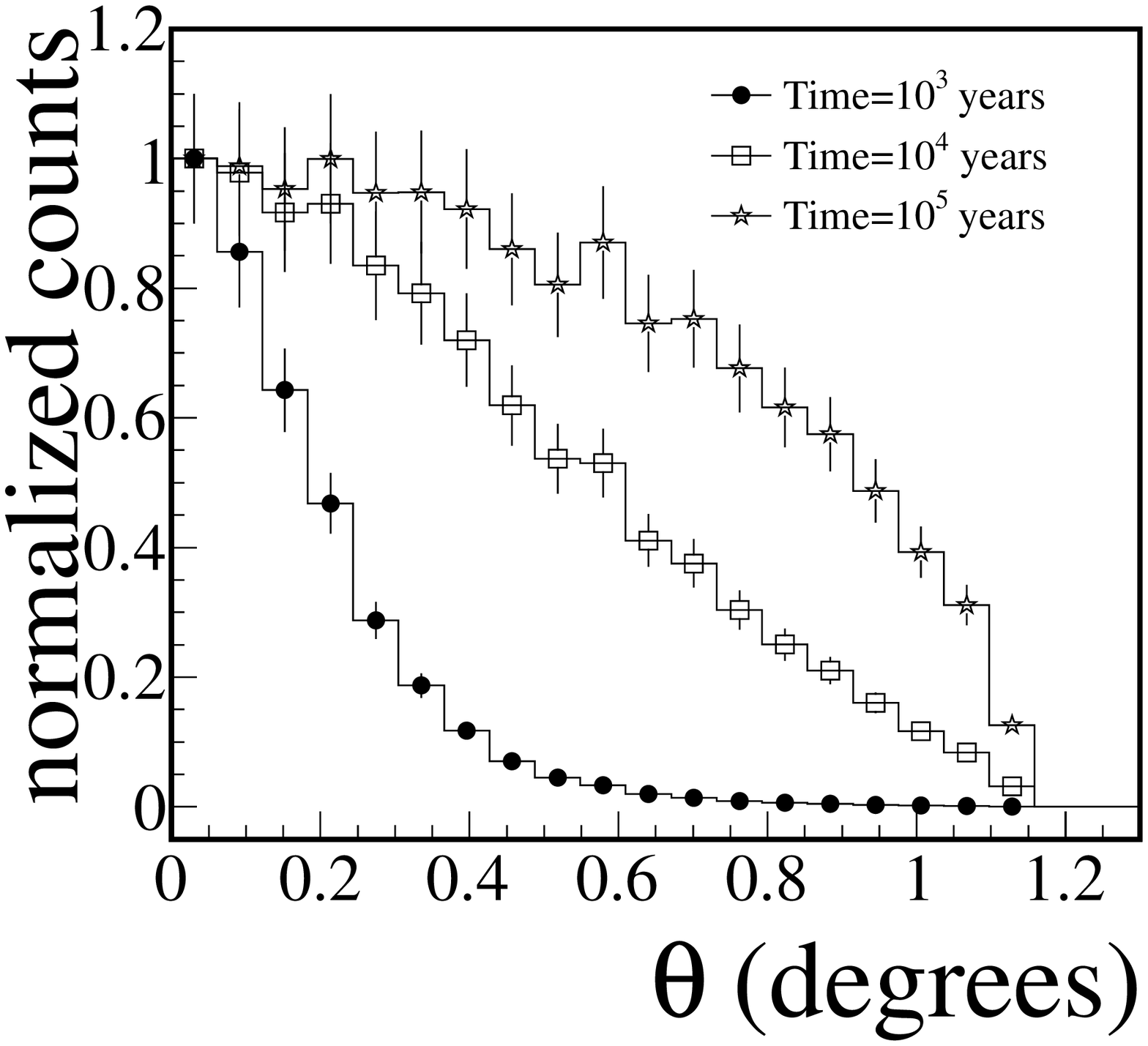}
\includegraphics[width=0.45\textwidth]{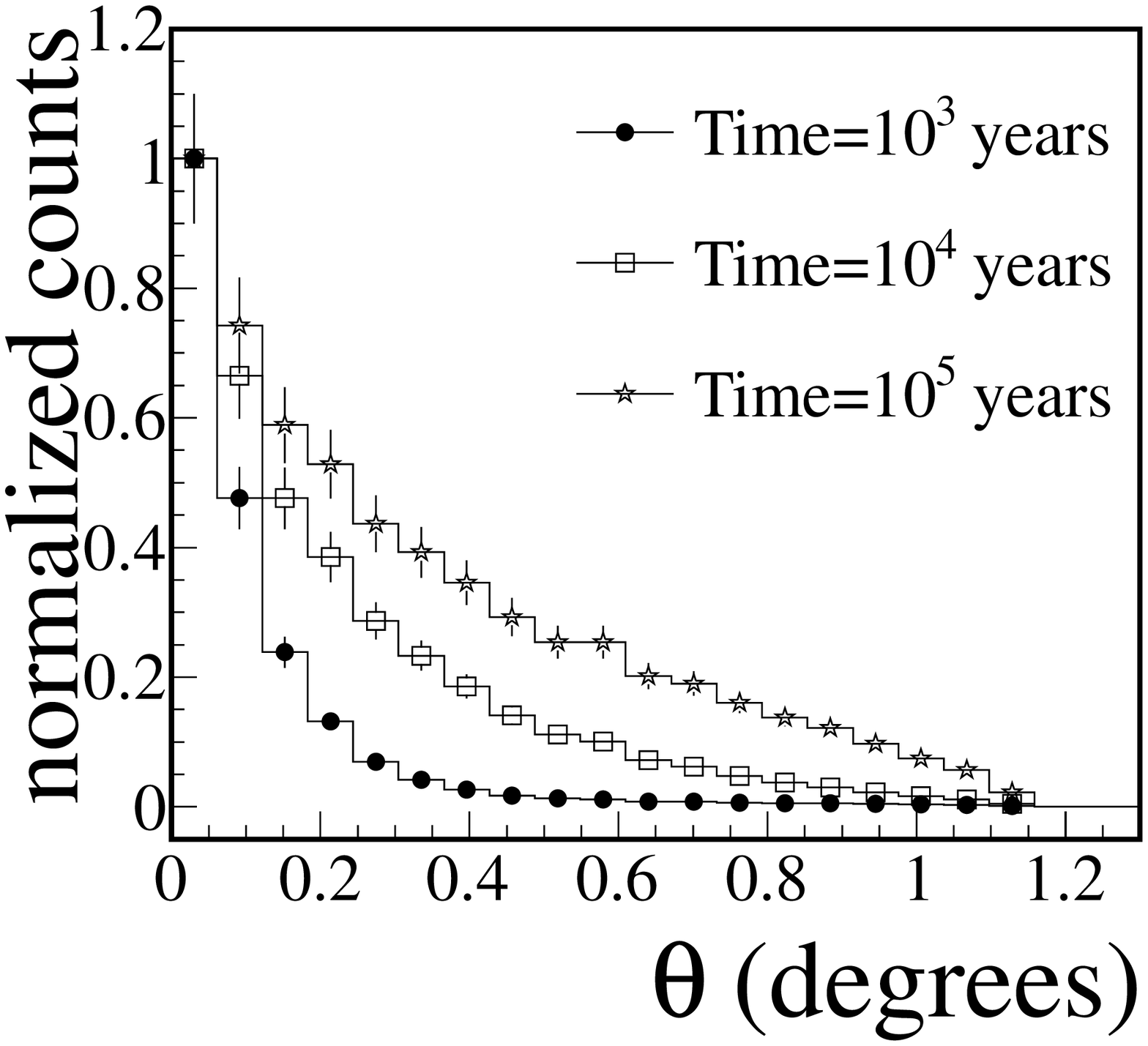}
\caption{Profiles of the photon count, normalized to the respective maximum. Here we show the example of impulsive (\textbf{left}) and continuous (\textbf{right}) acceleration, $D_{10}=10^{26}$ cm$^2$ s$^{-1}$, $\delta=0.5$, $\gamma=2.2$, $\eta=1$, at an accelerator age of $10^{3}$ (filled circles), $10^{4}$ years (open squares), and $10^{5}$ years (stars). Error bars are set to 10\% of the count number, to mimic the expected error on the effective area for the array used in the counts determination.}
 \label{fig:profiles}
\end{figure}
The different scenarios can also be disentangled by investigating the morphology and extension of the emission region.
The shape of those profiles depends on the parameters of the simulated scenario, an example of which is shown in Fig. \ref{fig:profiles}. These profiles are easily distinguishable from each other. Extensions of the \gr emission depend on the parameters studied here, with some general trends. Older sources are always more extended than younger sources, as the lower energy particles will have diffused further from the center of the cloud and thus from the accelerator. Emission due to continuous accelerators will present steeper profiles due to the freshly accelerated particles in the center of the source. Faster diffusion also leads to a larger extension. The profiles in Fig. \ref{fig:profiles} are normalized to their respective maximum counts, with flux decreasing with age in the impulsive case and opposite behavior in the continuous case. 
\\

{\footnotesize \textbf{Acknowledgements} We acknowledge discussions with Ana Y. Rodriguez-Marrero and many colleagues of the CTA collaboration, especially Sabrina Casanova, Elsa de Cea del Pozo, Daniela Hadasch, Matthieu Renaud, Jim Hinton, Konrad Bernloehr, and Abelardo Moralejo.
We acknowledge support from the Ministry of Science and the Generalitat de Catalunya, through 
the grants AYA2009-07391 and SGR2009-811, as well as by
ASPERA-EU through grant EUI-2009-04072. }


\begin{thebibliography}{99}

\bibitem[{Albert}(2007)]{ic443_magic}
{Albert} J. et al. 2007 ApJ, 664L, 87A

\bibitem[{Aharonian et al}(2008)]{w28_hess}
{Aharonian} F. et al. 20008 A\&A, 481, 401

\bibitem[{Ackermann et al.}(2012)]{mc_fermi}
Ackermann, M. et al. 2012 ApJ, 755, 22A

\bibitem[{Montmerle}(1979)]{montmerle79}
{Montmerle}, T. 1979 ApJ, 231, 95M

 \bibitem[{Aharonian \& Atoyan}(1996)]{aha_ato} 
{Aharonian} \& {Atoyan} 1996 ApJ, {309}, 917

\bibitem[{Gabici et al.}(2007)]{gabici2007} Gabici et al. 
		2007 AsSS, {309}, 365

\bibitem[{Rodr\'iguez Marrero et al.}(2008)]{rodriguez2008}
Rodr\'iguez Marrero A. Y. et al., 2008, ApJ, 689, 213

\bibitem[{Pedaletti et al.}(2012)]{longer}
Pedaletti et al., 2012, intended for A\&A

\bibitem[{Ginzburg \& Syrovatskii}(1964)]{ginzburg1964} 
Ginzburg V. L., Syrovatskii S. I., 1964, The Origin of Cosmic Rays. Pergamon Press, London

\bibitem[{Berezinsky}(1990)]{ber_book}
Berezinsky, V. S. et al.: Astrophysics of Cosmic Rays,
North Holland (1990)

\bibitem[{Tam et al.}(2010)]{tam_gevtev}
Tam, P. H. T., Wagner, S. J., Tibolla, O. \& Chaves, R. C. G. 2010 A\&A, 518A, 8T

\bibitem[{Actis et al.}(2011)]{dc2010}
{Actis}, M. et al. (CTA Consortium)  2011 ExA, 32, 193A

 \bibitem[{Atwood et al}(2009)]{atwood2009} 
Atwood, W. B., et al. 2009, ApJ, 697, 1071

\bibitem[{Goodman}(2010)]{hawc2010}
Goodman J. A., 	for the HAWC Collaboration 	
	2010 ASPC, 426, 19G

\bibitem[{Bernl\"{o}hr et al.}(2012)]{apissue_mc}	
Bernl\"{o}hr, K. et al. (CTA Consortium) 2012 Astroparticle Physics CTA Special Issue, in press

\end{thebibliography}
\end{document}